\documentstyle[12pt,epsfig]{article}

\def\beq{\begin{equation}}
\def\eeq{\end{equation}}
\def\beqa{\begin{eqnarray}}
\def\eeqa{\end{eqnarray}}

\begin{document}
\begin{titlepage}
\begin{flushright}    
		     RRI--96--14; gr-qc/9606020\\
\end{flushright}
\begin{center}
   \vskip 3em 
   {\LARGE  Black hole area in Brans-Dicke theory } 
   \vskip 1.5em
   {\large Gungwon Kang\footnote{kang@rri.ernet.in} 
   \\[.3em]}
{\em Raman Research Institute, Bangalore 560 080, 
India}  \\[.7em]  
\end{center}
\vskip 1em
\begin{abstract}

We have shown that the dynamics of the scalar field $\phi (x)=
``G^{-1}(x)"$ in Brans-Dicke theories of gravity makes the surface 
area of
the black hole horizon {\it oscillatory} during its dynamical
evolution. It explicitly explains why the area theorem does not 
hold in
Brans-Dicke theory. However, we show that there exists
a certain non-decreasing quantity defined on the event horizon 
which is proportional to the black hole entropy for the case of 
stationary solutions in Brans-Dicke theory. Some numerical 
simulations have been demonstrated for Oppenheimer-Snyder 
collapse in Brans-Dicke theory.  
\vspace{1cm}
\newline PACS number(s): 04.70.-s, 04.70.Bw, 04.70.Dy
   
\end{abstract}
\end{titlepage}
\newpage
\section{Introduction }
  
Scheel, Shapiro and Teukolsky\cite{SST} have numerically shown that
Oppenheimer-Snyder collapse in Brans-Dicke theories of gravity
produces black holes that are identical to those of general 
relativity
in the final stationary stage, but behave quite differently during
dynamical evolution. For example, in general relativity the apparent
horizon of a black hole is always inside the event horizon and the
total surface area of the event horizon never decreases in {\it any}
classical process provided that null energy condition for matter
fields(i.e., $T_{ab}k^ak^b \geq 0$ for all null vectors $k^a$) and
cosmic censorship conjecture are satisfied\cite{Areathm}.  In
Brans-Dicke theories of gravity, however, they found that there are
some initial epochs in Oppenheimer-Snyder collapse during which not
only the apparent horizon passes outside the event horizon, but also
the surface area of the event horizon decreases in time. Thus, the
area increase theorem for black holes in general relativity does not
hold in Brans-Dicke theory.  As mentioned briefly in Ref.~\cite{SST},
these different behaviors are possible because the null convergence
condition, i.e., $R_{ab}k^ak^b \geq 0$ for all null vectors $k^a$, is
violated for a dynamical black hole in Brans-Dicke theory.  At the
present paper, we investigate how the behavior of the auxilliary
scalar field $\phi (x)=``G^{-1}(x)"$ in Brans-Dicke theory causes 
this
violation and the oscillatory behavior of the surface area in detail.

One may think that the violation of the area theorem for black holes
in Brans-Dicke (BD) theory causes some problem to the black hole 
thermodynamics
in BD gravity since the black hole entropy in Einstein gravity is 
propotional to the surface area of the horizon and so the area 
theorem automatically serves as the classical second law in black 
hole mechanics. We show that there 
exists a certain quantity defined on the event horizon which never 
decreases during any classical process in BD gravity. Moreover, 
this quantity coincides with the entropy in the case of stationary 
black holes in BD theory. Thus, the black hole thermodynamic second 
law in BD theory is established. 
Since BD theories of gravity are dynamically related to other 
theories of gravity such as higher curvature theories of gravity
and dilaton gravity, 
the results for black holes in BD theory also give 
some insights for those in other theories of gravity. 

In sec.~\ref{BHABDT}, we explain why black holes in BD theory 
behave differently from those in Einstein gravity by investigating
the behavior of the auxilliary scalar field $\phi (x)$. 
In sec.~\ref{AQAI}, a quantity is constructed on the event horizon
from various points of view and proved that it is always 
non-decreasing for arbitary dynamical processes. 
Finally, the validity of assumptions on the scalar field $\phi (x)$
used in the proof is discussed.  
General feature of the proof and possible application to other 
theorems are also mentioned briefly. 
 
\section{Black hole area in Brans-Dicke theory}
\label{BHABDT}

Let us consider the change of the total surface area of the horizon 
of any black hole along the null congruence of the horizon generators
orthogonal to the spacelike cross-section ${\cal H}$:
\beq
\frac{dA(\lambda )}{d\lambda }=\frac{d}{d\lambda }\oint_{\cal H}
d^2x\sqrt{h} =\oint_{\cal H}d^2x\sqrt{h}\theta (\lambda )
\label{darea}
\eeq
where $\lambda$ is the affine parameter of null geodesics whose
tangents are $k^a$, i.e., $k^a\nabla_a=d/{d\lambda }$, and 
$\theta =d(\ln \sqrt{h})/{d\lambda}=\nabla_ak^a$
is the expansion of the horizon generators. 
One can easily see that, if $\theta (\lambda )\geq 0$ 
everywhere on the horizon and at any point in $\lambda$, 
$dA/{d\lambda }\geq 0$ and so the surface area is non-decreasing
always. Otherwise, it either still increases or decreases
depending on the value of integration of $\theta$ over the whole 
event horizon\cite{footnote0}. 

\begin{figure}[btp]
\epsfysize=4.5cm
\hspace{3.5cm}
\epsfig{file=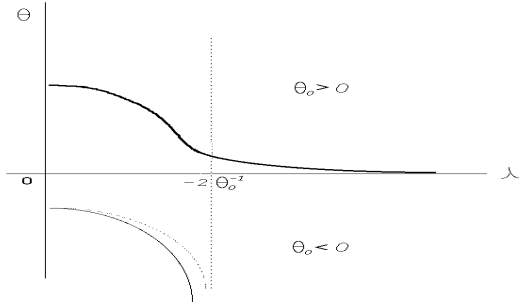,width=10cm,height=6cm}
\caption{Behavior of the expansion $\theta$
         in general relativity.}
\label{thetaGR}
\end{figure}

Now let us consider the evolution of expansion along a hypersurface 
orthogonal null congruence of geodesics in general. It is 
determined by Raychaudhuri equation 
\beq
\frac{d\theta }{d\lambda }=-\frac{1}{2}\theta^2-\sigma^2-R_{ab}k^ak^b
\label{Ray}
\eeq
where $\sigma^2$ is the square of the shear. 
In the case that $R_{ab}k^ak^b\geq 0$ for any null vector $k^a$,
$d\theta /{d\lambda }$ is negative-definite and
\beq
\frac{d\theta }{d\lambda }\leq -\frac{1}{2}\theta^2.
\nonumber 
\eeq
Or
\beq
\frac{d\theta^{-1}}{d\lambda } \geq \frac{1}{2}. 
\eeq
Suppose one has $\theta < 0$ at some point on a null geodesic.  
Then the above inequality equation tells that $\theta^{-1}$ 
should meet zero within some finite affine parameter. That is to 
say, $\theta$ reaches $-\infty$ at some finite affine parameter,
giving a conjugate point as can be seen in Fig.~\ref{thetaGR}. 

When applied to black holes in a strongly asymptotically predictable
spacetime(thus, no naked singularity exists outside the black hole
region), this behavior of expansion leads that $\theta$ cannot be
negative, i.e., $\theta \geq 0$ everywhere on the horizon and at any
point in $\lambda$\cite{Areathm}; a similar derivation will be
explained more detaily in sec.~\ref{AQAI}. It, therefore, shows
$dA/{d\lambda }\geq 0$ proving the area theorem: {\it the total
surface area never decreases in any classical process.}

\begin{figure}[tb]
\epsfysize=4.5cm
\hspace{2cm}
\epsfig{file=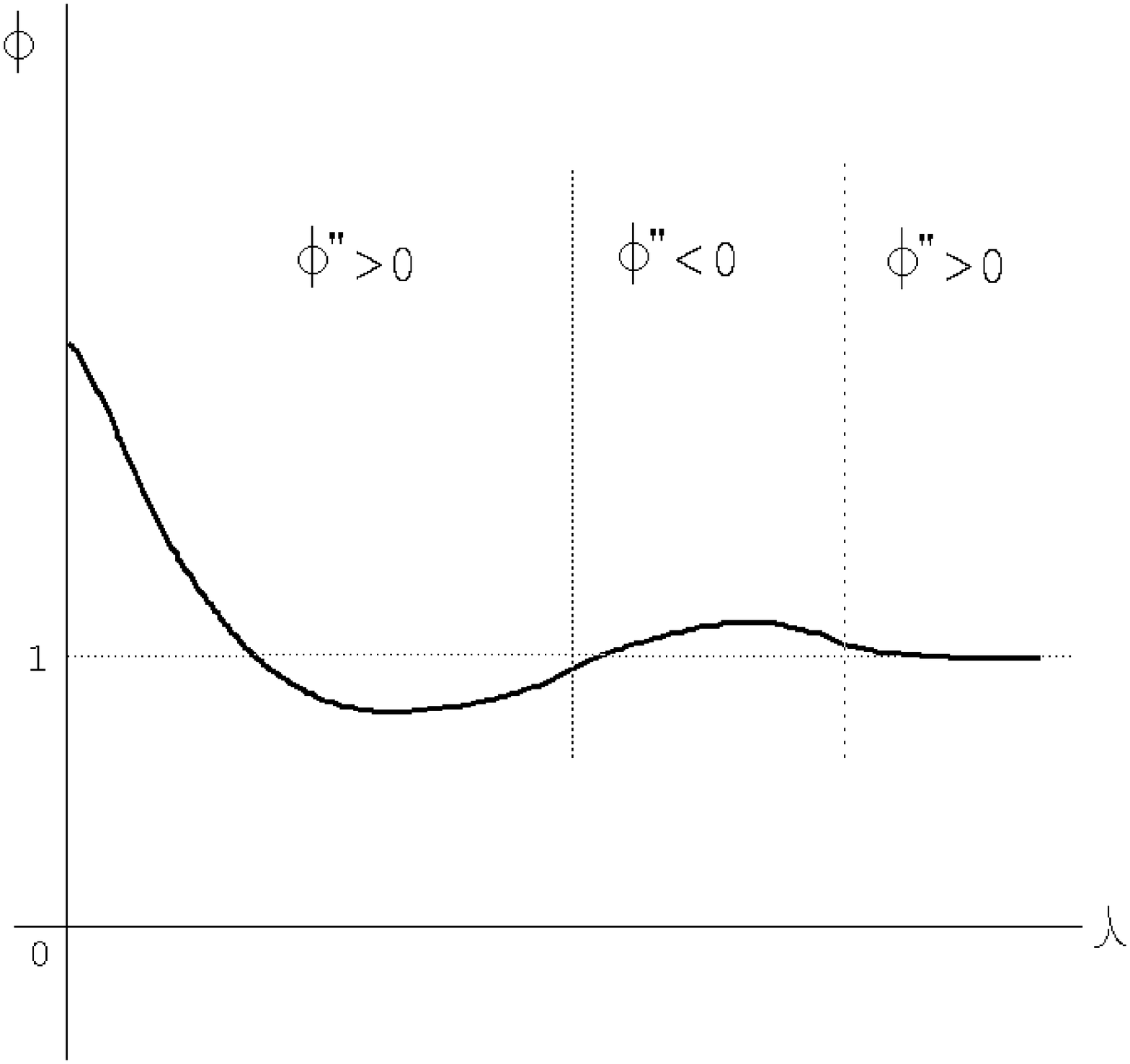,width=10cm,height=6cm}
\caption{Oscillatory behavior of the scalar
         field $\phi (x)$ in Brans-Dicke theory.}
\vspace{1cm}
\label{phi}
\end{figure}

\begin{figure}[tb]
\epsfysize=4.5cm
\hspace{2cm}
\epsfig{file=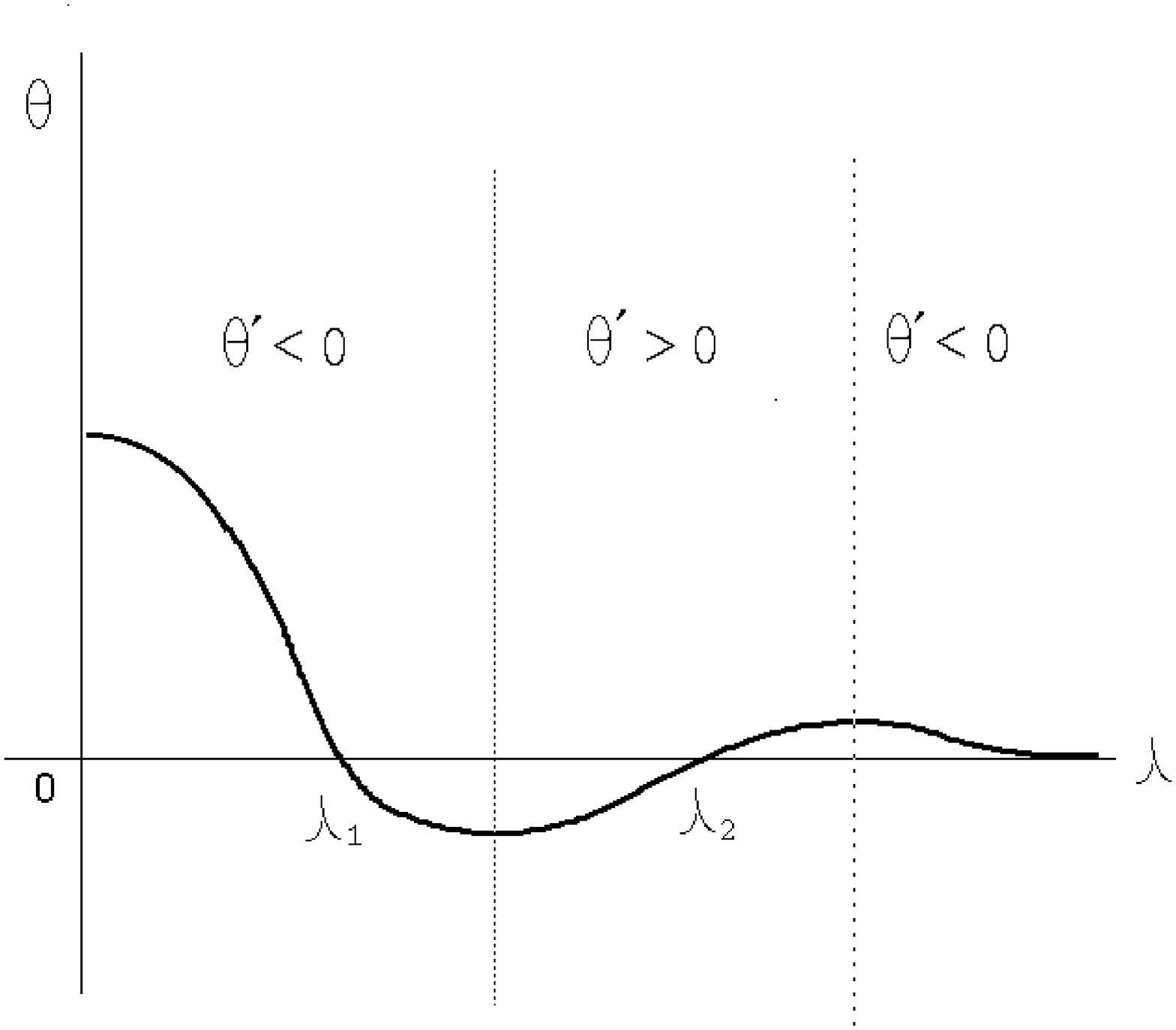,width=10cm,height=6cm}
\caption{ Possible behavior of the expansion 
         $\theta$ in Brans-Dicke theory corresponding to 
         that of $\phi$ in Fig.~\ref{phi}.}
\label{thetaBD}
\vspace{1cm} 
\end{figure}

For black holes in Einstein gravity, the null 
convergence condition is 
equivalent to the null energy condition since field equations are 
\beq
R_{ab}-\frac{1}{2}Rg_{ab}=8\pi GT_{ab}. 
\nonumber 
\eeq
For black holes in BD gravity, however, $R_{ab}k^ak^b$ becomes {\it 
indefinite} as follows;   
\beq
{\cal L}_{\rm BD} = \frac{1}{16\pi G}(R\phi -\frac{\omega }{\phi }
\nabla_a\phi \nabla^a\phi )+{\cal L}_{\rm matter}.
\eeq
Field equations are 
\beq
(R_{ab}-\frac{1}{2}Rg_{ab})\phi =8\pi GT_{ab}+\frac{\omega }{\phi }
(\nabla_a\phi \nabla_b\phi -\frac{1}{2}g_{ab}\nabla_c\phi 
\nabla^c\phi )
+\nabla_a\nabla_b \phi -g_{ab}\nabla_c\nabla^c\phi .
\label{FEQBD}
\eeq
And so
\beq
R_{ab}k^ak^b= 8\pi G\phi^{-1}T_{ab}k^ak^b+\omega \phi^{-2}
(k^a\nabla_a\phi )^2+\phi^{-1}k^ak^b\nabla_a\nabla_b\phi .
\label{CFEQBD}
\eeq
One sees that, even if null energy condition and positivity of $\phi$ 
and $\omega$ are satisfied, $R_{ab}k^ak^b$ is {\it indefinite}  
since the last term, $\phi^{-1}k^ak^b\nabla_a\nabla_b\phi =\phi^{-1}
k^a\nabla_a(k^b\nabla_b\phi )=\phi^{-1}d^2\phi /{d\lambda^2}=\phi^{-1}
\phi^{\prime \, \prime }$, could be strongly negative. Now  
Eq.~(\ref{Ray}) becomes
\beq
\theta'=-[\frac{1}{2}\theta^2+\sigma^2+8\pi G\phi^{-1}T_{ab}k^ak^b
+\omega (\phi'/\phi )^2]-\phi^{-1}\phi''. 
\label{RayBD}
\eeq
Therefore, if the auxilliary scalar field $\phi (x)$ behaves as in 
Fig.~\ref{phi} 
during the dynamical evolution 
of black holes, the 
expansion $\theta (\lambda )$ possibly behaves 
as in Fig.~\ref{thetaBD}
along the horizon generators. 
In other words, before $\theta$ hits the 
negative infinity after $\lambda_1$ in Fig.~\ref{thetaBD}, the 
negativeness of $\phi''$
starts to increase $\theta$ up over zero.  
Consequently, the total surface area of the horizon 
is increasing until $\lambda =\lambda_1$, 
decreasing between $\lambda_1$ 
and $\lambda_2$, increasing again after $\lambda =\lambda_2$, and  
finally approaches constant. 

For Oppenheimer-Snyder collapse in BD theory where an initially
uniform particle distribution within a sphere is allowed to collapse
to a final state, the numerical results obtained in Ref.~\cite{SST}
demonstrate this behavior described above. That is, Figure 10
in Ref.~\cite{SST} shows the indefiniteness of
$R_{ab}k^ak^b$ for different values of $\omega$, and Figure 9 in
Ref.~\cite{SST} shows that the surface area is strongly oscillating
for small $\omega$ and becomes monotonically increasing for large
$\omega$.  A monotonic behavior for large $\omega$ appears because,
even though BD theory does not completely reduce to Einstein gravity
in general as $\omega \rightarrow \infty$, it does for physical cases
as pointed out in Ref.~\cite{SST}. Figure 9 in Ref.~\cite{SST}
also shows that apparent horizons pass over the event
horizons during epochs of $\theta \leq 0$. It is simply because the
expansion of null geodesics is negative 
on the event horizon for those
periods. Then trapped surfaces can exist outside the event horizon in
contrary to black hole solutions in general relativity. The outer
boundary of this trapped region where expansion vanishes will be the
apparent horizon. Interestingly, ``outgoing" null geodesics from a
trapped surface outside the event horizon do not hit the
singularity. Instead they again expand afterwhile perhaps due to the
last term in Eq.~(\ref{RayBD}) and escape to the null infinity
finally.

 
\section{A non decreasing quantity}
\label{AQAI}

In the thermodynamic analogy of classical black hole mechanics in 
Einstein gravity, the total surface area of the event horizon plays 
a somewhat special role. 
It is proportional to the black hole entropy, 
which has been later justified by considering 
quantum effect on matter 
fields around the black hole. 
Hawking's area increase theorem then establishes the 
classical second law of black hole mechanics in Einstein gravity. 
In the previous section, however, 
we have seen that the area theorem does 
not hold for black holes in Brans-Dicke theory. 
Does this mean that the 
second law does not hold for black holes in BD 
theory? This question turns out to be irrelevent since in general  
``area-entropy" relation no longer holds in other 
theories of gravity\cite{areabhe}. Indeed this modification of 
``area-entropy" relation happens for black holes in BD theory. 

The entropy for stationary black holes 
in Einstein gravity is given by 
\beq
S_{\rm E}=A/{4G}=\frac{1}{4G}\oint_{\cal H}d^2x\sqrt{h}.
\label{bheGR}
\eeq
Since Newton's constant $G$ becomes dynamical in BD theory, 
a natural candidate for the entropy 
of stationary black holes in BD theory will be 
\beq
S_{\rm BD}=\frac{1}{4G}\oint_{\cal H}d^2x\sqrt{h}\phi (x).
\label{bheBD}
\eeq
Note that we defined $\phi (x)$ to be dimensionless. This quantity
is indeed the black hole entropy 
for stationary solutions in BD theory 
evaluated by many other 
methods such as Noether charge method, 
field redefinition, and Euclidean 
method\cite{onS}. 

The formula above can also be obtained by considering a ``physical
process" derivation of the first law in black hole thermodynamics
introduced by Wald\cite{Waldbook} and extended to a certain class of
higher curvature theories in Ref.~\cite{JKM}. Consider a black hole
stationary initially, being perturbed by small amount of matter
falling, and finally settling down to a stationary state again. Small
amount of matter falling into a stationary black hole, $\Delta
M=\int_OT_{ab}\xi^ad\Sigma^b$ and $\Delta
J=-\int_OT_{ab}\varphi^ad\Sigma^b$ in the asymptotically flat region,
produces some change at the event horizon as follows \beq \Delta
M-\Omega \Delta J= \int_OT_{ab}\chi^ad\Sigma^b
=-\int_HT_{ab}\chi^ad\Sigma^b=\kappa \int_HT_{ab}k^ak^b\lambda
\sqrt{h}d\lambda d^2x \eeq where $\chi^a=\xi^a+\Omega \varphi^a$ is
the Killing generator of the horizon in leading order of the
perturbation and $H={\cal H}\otimes {\bf R}$ the black hole horizon;
see details in Ref.~\cite{JKM}.  From field equations in (\ref{FEQBD})
one obtains in leading order \beqa \int_Hd^2xd\lambda \sqrt{h}\lambda
T_{ab}k^ak^b &=& \frac{1}{8\pi G}\int_{\cal
H}d^2x\int^{\lambda_f}_{\lambda_i}d\lambda \sqrt{h}\lambda (\phi
R_{ab}k^ak^b -\frac{\omega}{\phi}{\phi'}^2-\phi'') \nonumber \\
&\simeq & \frac{1}{8\pi G}\int_{\cal H}d^2x
\int^{\lambda_f}_{\lambda_i}d\lambda \sqrt{h}\lambda (-\phi
\theta'-\phi'') \nonumber \\ &\simeq & \frac{1}{8\pi G}\int_{\cal
H}d^2x[(\sqrt{h}\phi )\, |^{\lambda_f}_{\lambda_i}-\sqrt{h}\lambda
(\phi \theta +\phi')\, |^{\lambda_f}_{\lambda_i}].  \eeqa Since
$\theta =\phi'=0$ on both initial and final stationary stages, one
finds \beq \Delta M-\Omega \Delta J\simeq \frac{\kappa}{8\pi G}
\int_{\cal H}d^2x(\sqrt{h}\phi )\, |^{\lambda_f}_{\lambda_i}
=\frac{\kappa}{2\pi}\Delta S_{\rm BD}, \eeq giving the formula in
(\ref{bheBD}) for black hole entropy in BD theory.

Now let us assume that the quantity in (\ref{bheBD}) 
is the black hole 
entropy even at any moment of dynamical evolution of the horizon and 
see how it behaves.
To be more general in the proof, let us consider the 
following quantity
\beq
S=\frac{1}{4G}\oint_{\cal H}d^2x\sqrt{h}e^{\rho}
\label{bheG}
\eeq
where $e^{\rho}$ is a scalar function locally defined on the 
horizon. As before, the change of $S(\lambda )$ 
along the null congruence 
generating the event horizon is 
\beq
\frac{dS}{d\lambda}=\frac{1}{4G}\oint_{\cal H}d^2x\sqrt{h}e^{\rho}
\tilde{\theta}       \nonumber
\eeq
with
\beq
\tilde{\theta} = \theta +\partial_{\lambda}\rho .
\eeq
Now let us apply the method developed in Ref.~\cite{JKM}
which basically follows Hawking's proof of the area theorem, 
with $\tilde{\theta}$ in place of $\theta$.
Again the question is whether or not there can exist 
a point along the 
null geodesics at which $\tilde{\theta}$ becomes negative. 
The Raychaudhuri equation shows 
\beq
\partial_{\lambda}\tilde{\theta}=
-\frac{1}{2}\theta^2-\sigma^2-\omega^2
-(R_{ab}-\nabla_a\nabla_b\rho )k^ak^b.
\eeq
If the last term is positive-definite, then one has 
\beq
\partial_{\lambda}\tilde{\theta}\leq -\frac{1}{2}\theta^2,
\eeq
or
\beq
\partial_{\lambda}{\tilde{\theta}}^{-1} \geq \frac{1}{2}(\theta 
/{\tilde{\theta}})^2.
\label{Rayineq}
\eeq 
Now suppose $\tilde{\theta}<0$ at some point on the horizon.
Then in a neighborhood of that point one can deform a spacelike slice
of the horizon slightly outward to obtain a compact spacelike surface
$\Sigma$ which enters $J^-({\cal I}^+)$ and has $\tilde{\theta}<0$
everywhere on $\Sigma$, $\tilde{\theta}$ being defined along the
outgoing null geodesic congruence orthogonal to $\Sigma$. If cosmic
censorship is assumed, then there is necessarily some null geodesic
orthogonal to $\Sigma$ that remains on the boundary of the future of
$\Sigma$ all the way out to ${\cal I}^+$\cite{Areathm}. In other
words, this geodesic has no conjugate point between $\Sigma$ and
${\cal I}^+$ and is future complete. However, this is impossible for
the following reason.  Asymptotic flatness implies that $\rho
\rightarrow 0$ like $\lambda^{-1}$ at infinity for the case of
(\ref{bheBD}) and so does $\theta$, where $\lambda$ is the affine
parameter along an outgoing null geodesic. Therefore $\theta
/{\tilde{\theta}}\rightarrow 1+O(\lambda^{-1})$, so the inequality
(\ref{Rayineq}) implies that, as one follows the geodesic outwards
from $\Sigma$, $\tilde{\theta}$ reaches $-\infty$ within some finite
affine parameter.  Since $\tilde{\theta}=\theta
+\partial_{\lambda}\rho $, this means that either $\theta$ or
$\partial_{\lambda}\rho$ goes to $-\infty$. In the former case we have
a contradiction, as in the area theorem, since it implies there is a
conjugate point on the geodesic.  In the latter case it leads
to a naked singularity outside the horizon, contradicting to cosmic
censorship, as shall be shown for BD theory below.

\begin{figure}[htb]
\epsfysize=4.5cm
\hspace{2cm}
\epsfig{file=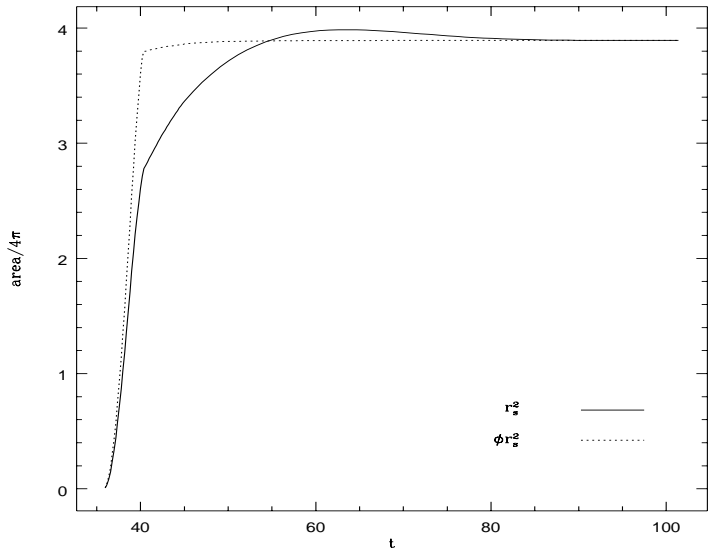,width=10cm,height=8cm}
\caption{Behavior of the surface area of the horizon 
         (solid line) and black hole entropy (dotted line) 
         in numerical simulation for Oppenheimer-Snyder collapse 
         in Brans-Dicke theory with $\omega =0$. Figure courtesy of 
         Mark A. Scheel and Richard O'Shaughnessy. The figure was 
         produced using the numerical code 
         described in PRD{\bf 51}, 4208 (1995).}  
\label{BHEBD}
\vspace{.5cm}
\end{figure}

\begin{figure}[tb]
\epsfysize=4.5cm
\hspace{2cm}
\epsfig{file=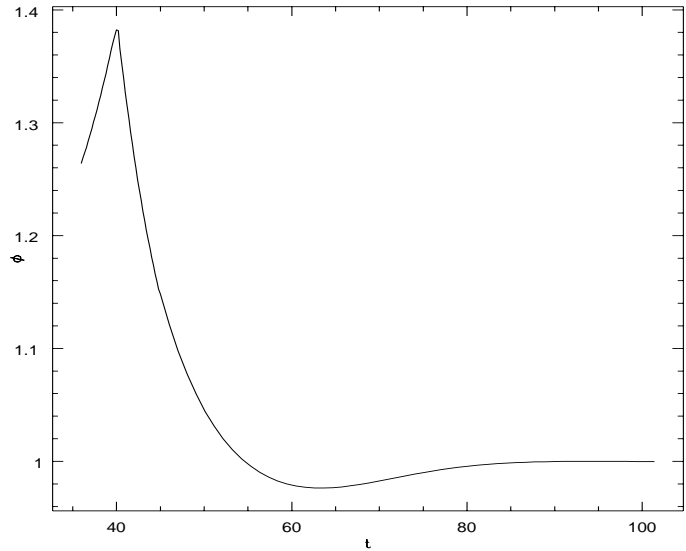,width=10cm,height=8cm}
\caption{Positivity and oscillatory behavior of
         the scalar field $\phi (x)$ corresponding to 
         Fig.~\ref{BHEBD}.
         Figure courtesy of Mark A. Scheel and Richard
         O'Shaughnessy. The figure was produced using 
         the numerical code
         described in PRD{\bf 51}, 4208 (1995).} 
\vspace{1cm}
\label{phiBD} 
\end{figure}

Therefore, one finally proves the entropy increase theorem for black 
holes in BD theory by checking whether or not the positivity of 
$(R_{ab}-\nabla_a\nabla_b\rho )k^ak^b$ is satisfied. Since 
$e^{\rho}=\phi (x)$, one has 
\beqa
(R_{ab}-\nabla_a\nabla_b\rho )k^ak^b &=& R_{ab}k^ak^b -k^ak^b\phi^{-1}
\nabla_a\nabla_b\phi +(k^a\nabla_a\phi /\phi )^2  \nonumber  \\
&=& 8\pi G\phi^{-1}T_{ab}k^ak^b +(1+\omega )(k^a\nabla_a\phi /\phi )^2
\label{GNC}
\eeqa 
which is manifestly non-negative provided that null energy
condition for matter fields and positivity of $\phi$ and $\omega$ are
satisfied.  Since $\partial_{\lambda}e^{\rho}=k^a\nabla_a\phi
=e^{\rho}\partial_{\lambda}\rho =\phi \partial_{\lambda}\rho $ and
$\phi$ is assumed to be positive($\neq 0$), the divergence of
$\partial_{\lambda}\rho$ implies a curvature singularity from the
equation (\ref{CFEQBD}) unless it is cancelled by other terms 
on the right hand side 
of the equation.  Note finally that the proof described above
does not require the existence of regular event horizon as in
Hawking's area theorem.  A numerical simulation for Oppenheimer-Snyder
collapse in BD theory with $\omega =0$ in Fig.~\ref{BHEBD}
demonstrates that the quantity defined in Eq.~(\ref{bheBD}) is always
increasing even if the surface area of the event horizon is
oscillatory.  From the behavior of the surface area of the horizon one
sees that $\theta$ is positive initially, zero near $t=60$, becomes
negative, and increases to zero finally. In order for $\theta$ to
increase from some negative value near the final stationary stage, the
only possible way is that $\phi''$ is negative for that period and
dominant in the equation (\ref{RayBD}). In fact $\phi''$ turns to be
negative near $t=75$ in Fig.~\ref{phiBD}. For other values of $\omega$
one can refer to Figure 12 in Ref.~\cite{SST}. The black hole area in
conformally related Einstein frame is indeed the same as black hole
entropy in BD theory as shall be shown in Eq.~(\ref{bhes}) below.

\section{Discussion}
\label{DISC}

In summary we have shown that the decrease of black hole area during 
the dynamical epoch of Oppenheimer-Snyder collapse 
in BD theory is due to 
the oscillatory behavior of the auxilliary scalar field 
$\phi (x)=``G^{-1}(x)"$ in strong gravitational field region. 
We also have 
proved that the quantity defined in (\ref{bheBD}), 
which is indeed the 
entropy of stationary black holes in BD theory, never decreases under 
arbitrary dynamical evolutions. Thus, it establishes a second law of 
black hole thermodynamics in BD theory. 

Results for dynamical black holes in BD theory also give some hints on
the behavior of black holes in other theories of gravity. 
For example, 
let us consider curvature scalar squared theories of gravity
\beq
I_0=\int d^4x\sqrt{-g}[\frac{1}{16\pi G}(R+\alpha R^2)+
    {\cal L}_m(\psi ,g)].
\label{R2}
\eeq
This higher curvature theory is dynamically equivalent 
to the following 
theory with no higher curvature term but with one auxilliary field 
$\Phi (x)$\cite{Wands} 
\beq
I_1=\int d^4x\sqrt{-g}[\frac{1}{16\pi G}(1+2\alpha \Phi )R-
    \frac{\alpha }{16\pi G}\Phi^2+{\cal L}_m(\psi,g)].
\label{MR2}
\eeq
By redefining the scalar field $1+2\alpha \Phi =\phi^q$ with 
$q^2=1+2\omega /3$ and conformally scaling the metric 
$g_{ab}=\phi^{1-q}{\bar{g}}_{ab}$, 
one finds that $I_1$ has the form of 
BD theory with a potential term and unconventional 
couplings between the 
scalar field $\phi$ and matter fields as follows
\beq
I_2=\int d^4x\sqrt{-\bar{g}}[\frac{1}{16\pi G}(\phi \bar{R}
    -\frac{\omega}{\phi}{\bar{\nabla}}_a\phi {\bar{\nabla}}^a\phi )
    -V(\phi )+\phi^{2(1-q)}{\cal L}_m(\psi ,\phi^{1-q}\bar{g})]
\label{BD}
\eeq
where $V(\phi )=\phi^2(1-\phi^{-q})^2/{16\pi G\alpha }$. 
Further conformally transforming 
${\bar{g}}_{ab}=\phi^{-1}{\tilde{g}}_{ab}$
and redefining the field $\varphi =\beta^{-1}\ln \phi $ with 
$\beta =\sqrt{16\pi G/(3+2\omega )}$, one reaches to the form of 
ordinary Einstein gravity\cite{Hawking,Wands} 
\beq
I_3=\int d^4x\sqrt{-\tilde{g}}[\frac{1}{16\pi G}\tilde{R}
    -\frac{1}{2}{\tilde{\nabla}}_a\varphi {\tilde{\nabla}}^a\varphi 
    -e^{-2\beta \varphi}V(\varphi )
    +e^{-2\beta q\varphi }{\cal L}_m(\psi ,
    e^{-\beta q\varphi }\tilde{g})].
\label{GR}
\eeq
Now let us ask how the black hole area 
in curvature squared theories behaves. 
First of all, notice that the null convergence condition 
is not satisfied 
in this theory as well. 
Thus, one expects that the black hole area will be 
oscillatory durinng dynamical epochs as in the case of BD theory. The
relationship between $I_0$ and $I_2$ shows it more explicitly. 
Assume an initial Cauchy surface such as that of Oppenheimer-Snyder
collapse in $I_0$ where spacetime is almost flat, e.g., $R\simeq 0$.
Since $\Phi =R$ in $I_1$, $\phi \simeq 1$ and $V(\phi )\simeq 0$ 
on the 
corresponding Cauchy surface in $I_2$. From the results for BD theory
above, one then expects that the behavior of $\phi$ is 
oscillatory during
dynamical epoch of the collapse. 
The ``extended" null convergence condition 
in (\ref{GNC}) to prove the 
increasement of $\sqrt{\bar{h}}\phi$ is still
satisfied in $I_2$ even in the presence of 
the potential $V(\phi )$ and 
the unconventional couplings between $\phi (x)$ 
and matter fields as long 
as $\phi$ and $\omega$ are positive. 
Thus the quantity $\sqrt{\bar{h}}\phi =\sqrt{h}\phi^q$ 
on the event horizon
never decreases. 
Then the ``area" element $\sqrt{h}=(\sqrt{\bar{h}}\phi )/
{\phi^q}$ will be in general oscillatory 
for black holes in $I_0$ unless 
the ratial increasement of $\sqrt{\bar{h}}\phi$ is 
bigger than that of 
$\phi^q$ for the increasing period. 
However, the positivity of $\phi (x)$
is guaranteed only for $\alpha >0$. 
It follows because the form of $V(\phi )$ 
confines the field $\phi$ in the positive region 
only if $\alpha >0$ since
the potential barrier increases exponentially as $\varphi \rightarrow 
-\infty$ in $I_3$(i.e., $\phi \rightarrow 0$ in $I_2$). 
When $\alpha <0$, the potential falls down exponentially, 
$\phi$ easily becomes negative, and then the conformal factors
become singular at $\phi =0$. 
It also has been shown that the theory $I_0$ has
the well-posed initial value formulation only if 
$\alpha >0$\cite{CauchyR2}. 

As byproducts of the relationships above one finds 
\beqa
{\tilde{S}}_{\rm BH}=\frac{\tilde{A}}{4G} &=& \frac{1}{4G}
	\oint_{{\cal H}_3}d^2x\sqrt{\tilde{h}}=\frac{1}{4G}
	\oint_{{\cal H}_2}d^2x\sqrt{\bar{h}}\phi    \nonumber  \\
&=& \frac{1}{4G}\oint_{{\cal H}_1}d^2x\sqrt{h}(1+2\alpha \Phi )
 =  \frac{1}{4G}\oint_{{\cal H}_0}d^2x\sqrt{h}(1+2\alpha R).
\label{bhes}
\eeqa
Therefore, one sees that Hawking's area theorem is transfered to 
the increasement of black hole entropy in BD theory 
as well as in higher
curvature theories through conformal transformations 
and an introduction
of a scalar field\cite{footnote1}. 
The anaysis above can also be extended
to more general class of higher curvature theories such as 
actions polynomial in $R$\cite{JKM}. 

In the proof given in sec.~\ref{AQAI}, we assumed the positivity of
the scalar field $\phi (x)$ and the coupling constant $\omega$ in BD
theory.  The positivity of $\omega$ is natural since, otherwise, it
gives unphysical negative energy matter in the theory. However, the
positivity of $\phi (x)$ at any moment is highly non-trivial in BD
theory and should be guaranteed in order for the quantity in
(\ref{bheBD}) to be interpreted as ``entropy," which is positive by
definition in statistical mechanics.  From field equations in
(\ref{FEQBD}) we see that any constant $\phi$ can be a solution. Thus,
each set of solutions for a given $\phi = {\rm const.}$(which could be
negative as well) in BD theory with large $\omega$ gives Einstein-like
gravity. For Einstein gravity this constant is determined by
considering Newtonian limit. It gives a positive constant $\phi =1$ in
the unit of $G^{-1}$ and so the black hole entropy (\ref{bheGR}) in
Einstein gravity is always positive. In BD theory, however, $\phi (x)$
is highly dynamical and so the quantity in (\ref{bheBD}) could be
negative at some stage, which is problematic to be an entropy of
dynamical black holes in BD theory. We are in fact interested in a
physical system in which the matter is initially distributed in an
almost flat space. In other words, $\phi (x)$ is initially positive
near the unit.  Now the question is whether or not there is a Cauchy
surface in the future evolution where the field $\phi (x)$ passes zero
and becomes negative.  As can be seen in field equations
(\ref{FEQBD}), it implies a curvature singularity unless the total
energy-momentum tensor on the right hand side cancels this singular
behavior. Therefore, $\phi \rightarrow 0$ corresponds to a naked
singularity on the event horizon or outside the horizon, violating the
assumption of cosmic censorship in BD theory which we used in the
proof.  To see whether the dynamics of the field $\phi$ can prevent
this singular behavior, let us examine the relationship between BD
theory $I_2$ and Einstein gravity $I_3$. In the theory $I_3$, $\phi
=0$ corresponds to $\varphi \rightarrow -\infty $. In general, the
dynamics of the field $\varphi$ is determined by the potentials coming
from couplings with matter Lagrangian as well as $V(\phi )$. If the
net effect of potentials gives rising up as $\varphi \rightarrow
-\infty$, the field $\varphi$ cannot reach to $-\infty$. One example
is the higher curvature squared theory with $\alpha >0$ through
conformal rescalings as explained above.  For BD theory without
potential terms, that is, $V=0$ and $q=1$ in $I_2$, the rising up of
potantial barrier as $\varphi \rightarrow -\infty$ is not guaranteed
in general. For example, suppose a typical matter field, ${\cal
L}_m(\psi,\bar{g})\sim -\frac{1}{2}{\bar{\nabla}}_a\psi
{\bar{\nabla}}^a\psi -V(\psi )$. Then, \beq e^{-2\beta \varphi}{\cal
L}_m(\psi,e^{-\beta \varphi}\tilde{g})\sim
-[\frac{1}{2}({\tilde{\nabla}}_a\psi {\tilde{\nabla}}^a\psi )e^{-\beta
\varphi}+e^{-2\beta \varphi}V(\psi )].  \eeq Since
${\tilde{\nabla}}_a\psi {\tilde{\nabla}}^a\psi$ could be negative in
general, the potential could fall exponentially, leading to $\varphi
\rightarrow -\infty$ and so $\phi \rightarrow 0$.  As far as we know,
however, we have not found any physical system in the literature which
shows vanishing of $\phi (x)$. Numerical simulation for the case of
Oppenheimer-Snyder collapse in Fig.~\ref{phiBD} also shows that the
scalar field $\phi (x)$ is always positive during its evolution.

Finally, let us extract some general feature of the proof shown in
sec.~\ref{AQAI}. One of the most important behaviors of expansion
$\theta$ used in proving many global properties of spacetime such as
singularity theorem and area theorem is that $\theta$ reaches
$-\infty$ within some finite affine parameter if $\theta$ is negative
at some point and convergence condition(i.e., $R_{ab}v^av^b \geq 0$
for all non-spacelike vectors $v^a$) is satisfied as explained in
sec.~\ref{BHABDT} briefly.  In general relativity the convergence
condition is satisfied if some suitable energy condition for matter
fields is assumed. In other theories of gravity, however, this is not
true any more as seen, for instance, in sec.~\ref{BHABDT}. Thus, it is
not clear whether or not some global theorems in general relativity or
modified forms of them where the convergent behavior of $\theta$ is
used for proof still hold in other theories of gravity. In this paper
we have shown that a modified ``expansion" $\tilde{\theta}$ plays the
same role as $\theta$ under same energy conditions for matter fields.
For example, one may define a modified ``trapped" surface as a
compact, two dimensional, smooth spacelike surface where all
``ingoing" and ``outgoing" null geodesics orthogonal to it have
negative $\tilde{\theta}$, instead of having negative $\theta$.  In
fact, for BD theory, this $\tilde{\theta}$ is nothing but the
expansion in the conformally related Einstein frame with an overal
multiflication factor. However, the general feature of the proof in
this paper may still be applicable to other class of gravity theories
which are not conformally related to Einstein gravity. In addition, it
also gives some hints on investigating how robust many global
properties of spacetime in general relativity are under the change of
dynamics of gravitational fields.  In other words, by using the
convergent behavior of $\tilde{\theta}$, we may extend the validity of
many global results in general relativity to other theories of gravity
as done for the area theorem in this paper.

\vskip 1cm
\noindent
The author would like to acknowledge useful 
discussions with J. Samuel, T. Jacobson, J.H. Cho, and J.H. Yoon. 
Especially, I would like to thank M.A. Scheel and R. O'Shaughnessy
for their comments and kindly making diagrams used in this 
paper. 



\begin{thebibliography}{99}
\bibitem
{footnote0} However, if the collapse is spherically symmetric 
        as in the case
	of Oppenheimer-Snyder collapse, 
        the behavior of total surface area 
	is strictly determined by that of 
        the differential area element. 
        Note also that Hawking's area theorem holds 
        for any areal element 
        on the horizon.
\bibitem
{footnote1} It will be necessary to check whether or not 
        those black holes 
	in different theories are indeed same 
        in the context of black hole 
	thermodynamics. For example, 
        one can ask if their horizons, ${\cal H}_i$, and surface 
        gravities indeed agree. 
        For the case of higher 
        curvature theories and 
	Einstein gravity, see details in Ref.~\cite{JKM}.
\end{thebibliography}
\end{document}